\documentclass[amsmath,amssymb,aps,prd,onecolumn]{revtex4-2}
\usepackage{hyperref}
\usepackage{graphicx}
\usepackage{dcolumn}
\usepackage{bm}
\usepackage{float}

\usepackage{multirow}
\usepackage{CJKutf8, here}
\usepackage{amsmath}
\usepackage{color}
\usepackage{natbib}

\usepackage{ulem}

\DeclareRobustCommand{\ysso}{\bgroup\markoverwith{\textcolor{magenta}{\rule[.5ex]{2pt}{0.4pt}}}\ULon}

\DeclareRobustCommand{\kuso}{\bgroup\markoverwith{\textcolor{blue}{\rule[.5ex]{2pt}{0.4pt}}}\ULon}

\begin{document}

\preprint{aaa}

\title{Floquet Instability of Photon-Axion Conversion}

\author{Yuuki Sugiyama$^a$}
\affiliation{$^a$ 
Institute for Solid State Physics, the University of Tokyo, Kashiwa, Chiba 277-8581, Japan}

\author{Kazushige Ueda$^{b,c}$}
\affiliation{$^b$ 
Physics Program, Graduate School of Advanced Science and Engineering, Hiroshima University, 1-3-1 Kagamiyama, Higashi-Hiroshima, Hiroshima 739-8526, Japan
\\ 
$^c$Yasuda Girls' Junior \textsl{\&} Senior High School, Hiroshima 730-0001, Japan 
}

\begin{abstract}
The discovery of axions would provide evidence for physics beyond the Standard Model.
In this work, we investigate Floquet instability in axion-photon conversion motivated by photons orbiting a black hole. 
We consider a simplified model with a background electromagnetic field that varies periodically in both space and time.
By performing a Lorentz transformation, the coupled axion-photon system is mapped onto a time-periodic linear system that can be analyzed using Floquet theorem.
We find unstable modes associated with resonant crossing between Floquet replicas.
The resulting complex Floquet exponents lead to exponentially growing modes, which we identify as a Floquet instability.
Then, we determine the parameter regions in which this instability appears in a perturbative analysis.
We further show that the Floquet instability enhances the axion-field amplitude generated from an initial photon field.
Our results demonstrate how a periodically varying electromagnetic coupling can give rise to Floquet instability and affect axion-photon conversion in a setting motivated by photon motion around a black hole.
\end{abstract}
\maketitle

\section{Introduction}
Despite the remarkable success of the Standard Model of particle physics, several fundamental questions such as the strong CP problem in quantum chromodynamics and the dark matter problem in cosmology remain unsolved, motivating searches for physics beyond the Standard Model.
As a possible solution to these problems, Axions and axion-like particles are often considered in the context of particle physics and cosmology.
The axion was originally introduced through the Peccei--Quinn mechanism as a dynamical solution to the strong $CP$ problem~\cite{PQ1977, PQ19772, Weinberg1978, Wilczek1978}.
A characteristic feature of these pseudoscalar particles is their coupling to the electromagnetic field, which provides an important basis for theoretical and experimental searches.
Even though extensive experimental searches over a wide range of axion parameters are performed, no axion has yet been detected, as illustrated by the current constraints on the axion parameter space~\cite{AxionLimits2020}.
It motivates the exploration of complementary approaches to axion searches beyond relying on its direct interaction with Standard Model particles.
We focus on the axion-photon conversion as a possible avenue for proving axion.

Axion-photon conversion is a theoretical phenomenon in which photons can convert into axions in the presence of a constant background magnetic field, and vice versa~\cite{Maiani1986, Raffelt1988}.
It is a topic that has been frequently studied in the context of axion exploration in recent years~\cite{Ringwald2024, Baryakhtar2025}.
A closely related conversion phenomenon is graviton-photon conversion~\cite{Gertsenshtein1962, Raffelt1988, Chen1995, Cillis1996, Ueda2022, Kanno2023, Kanno20232, TIkeda2025, Alessandro2026}, for which an instability (a parametric resonance) of photons coupled to background gravitational waves has recently been discussed in Refs.~\cite{Brandenberger2023, Delgado2023}.
Axion-photon conversion in the presence of dynamical axion backgrounds has also been investigated~\cite{Masaki2020,Jiao2025,Yao2026,Brandenberger2026}.
Beyond such dynamical axion backgrounds, axion-photon conversion under temporally or spatially periodic magnetic fields has also been studied~\cite{Arias2016,Seong2024}.
In Ref.~\cite{Arias2016}, the conversion was analyzed without invoking the eikonal approximation, and it was shown that the conversion probability grows quadratically with time, leading to  an enhancement compared with the case of a static background magnetic field.
The authors in Ref.~\cite{Seong2024} showed semi-analytically that a spatially periodic magnetic field can resonantly enhance axion-photon mixing.
More recently, Yao et al.~\cite{Yao20262} investigated the Floquet-Bloch sideband structure of the Maxwell-axion system in the presence of a periodic axion background.
They clarified how resonant mode coupling gives rise to instabilities. 
In contrast, the present work considers a periodically varying electromagnetic background without a background axion field. 
We investigate how the resulting periodic axion-photon coupling affects the conversion process in a setting motivated by photons orbiting a black hole.

In this work, we theoretically investigate Floquet instability in axion-photon conversion for photons orbiting a black hole. 
In our simplified model, the photon-axion system encounters a periodically varying electromagnetic background along its propagation path as shown in Fig.~\ref{fig:config}. 
By performing a Lorentz transformation, we obtain a spatially homogeneous, time-periodic electric field that induces a periodic axion-photon coupling. 
We analyze the resulting Floquet spectrum and find unstable modes associated with resonant crossings between Floquet replicas, consistent with the Floquet-Bloch picture discussed in Ref.~\cite{Yao20262}. 
We clarify the origin of the instability by analyzing the real and imaginary parts of the Floquet exponent and determine the parameter regions in which unstable modes appear.
We further show that the axion-field amplitude generated from an initial photon field is enhanced by the Floquet instability.
This paper is organized as follows.
In Sec.~II, we introduce a model describing the coupled dynamics of the axion and electromagnetic fields and derive their equations of motion.
By performing a Lorentz transformation and recasting the coupled field equations into a first-order form, we obtain a linear system with time-periodic coefficients, to which the Floquet theorem can be applied.
In Sec.~III, we analyze the Floquet spectrum and clarify the origin of the Floquet instability in our system.
We also identify the parameter regions in which unstable modes appear.
In Sec.~IV, we evaluate the axion-field amplitude generated from an initial photon field and show that it is enhanced by the Floquet instability.
Section~V is devoted to the conclusion.
In Appendix~A, we present a perturbative analysis of the Floquet Hamiltonian near the resonant crossings between Floquet replicas.
Throughout this paper, we use natural units, $c=\hbar=1$, and adopt
the metric convention
$\eta_{\mu\nu}=\mathrm{diag}(-1,+1,+1,+1)$
with the Minkowski spacetime coordinates
$x'^{\mu}=(t',\bm{x}')$.

\section{Model}
\begin{figure}[h]
 \centering
  \includegraphics[width=1.0\linewidth]{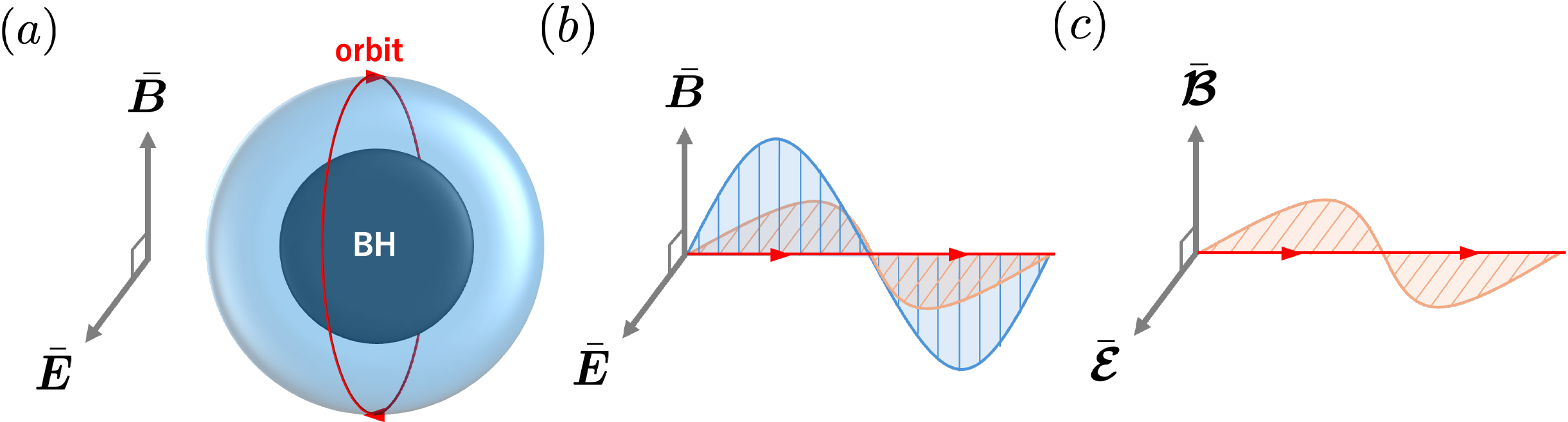}
   \caption
    {
    Schematic pictures of the axion-photon system and background electromagnetic field.
    (a) A photon undergoes circular motion along the photon sphere around a black hole immersed in a constant background electromagnetic field.
    (b) From the viewpoint of the orbiting photon, the constant background electromagnetic field is observed as a spatially and temporally periodic background electric and magnetic field, 
    $\bar{\bm{E}}$ and $\bar{\bm{B}}$, respectively.
    (c) By performing an appropriate Lorentz transformation, as given in Eq.~\eqref{Lorentz}, one can move to a frame in which the background magnetic field completely vanishes, 
    $\bar{\bm{\mathcal{B}}}=\bm{0}$, 
    while the electric field 
    $\bar{\bm{\mathcal{E}}}$ 
    oscillates periodically in time and remains spatially homogeneous.
    }
\label{fig:config}
\end{figure}

The conversion of photons orbiting a black hole into axions has previously been studied by Nomura et al.~\cite{Nomura2023}. 
In a similar setting, we investigate the possibility of Floquet instability and its effect on axion-photon conversion. 
We consider a simplified model in which photons propagate along a circular orbit around a stationary black hole in the presence of an external magnetic field, as illustrated in Fig.~\ref{fig:config}. 
The orbital motion motivates a periodically varying electromagnetic background along the propagation path, which we model by the prescribed field configuration introduced below. 
We focus on the resulting periodic axion-photon coupling and do not include gravitational effects in the equations of motion.

Here, we consider the coupled dynamics of an axion field and the electromagnetic field.
In the present paper, we use the convention $\epsilon_{0123}=+1$ for the totally anti-symmetric tensor
$\epsilon_{\mu\nu\alpha\beta}$.
We start from the following action
\begin{align}
S=\int d^4x'
\left[
-\frac{1}{4}F_{\mu\nu}F^{\mu\nu}
-\frac{1}{2}
\left(
\partial_{\mu'}a\partial^{\mu'}a+m^2_{a}a^2
\right)
-\frac{1}{4}
g_{a\gamma\gamma}aF_{\mu\nu}\tilde{F}^{\mu\nu}
-J^{\mu}A_{\mu}
\right],
\end{align}
where 
$F_{\mu\nu}=\partial_{\mu'}A_{\nu}-\partial_{\nu'}A_{\mu}$
with
$\partial_{\mu'}
:=\partial/\partial x'^{\mu}
=(\partial_{t'}, \bm{\nabla}')$
and
$A^{\mu}(t', \bm{x}')=(\phi(t', \bm{x}'), \bm{A}(t', \bm{x}'))$
is the electromagnetic tensor, and 
$\tilde{F}_{\mu\nu}=\frac{1}{2}\epsilon_{\mu\nu\alpha\beta}F^{\alpha\beta}$ 
is its dual.
The fields 
$\phi$
and
$\bm{A}$
are the scalar potential and the vector potentials, respectively.
The field $a$ is the axion field with mass $m_{\mathrm{a}}$, and $g_{a\gamma\gamma}$ 
denotes the axion-photon coupling constant.
The current $J^{\mu}$ represents an external source coupled to the electromagnetic field.
Varying the action with respect to 
$A_{\mu}(t', \bm{x}')$ and $a(t', \bm{x}')$, we obtain the following coupled-oscillator equations of motion
\begin{align}
\partial_{\mu'}F^{\mu\nu}
&=
-g_{a\gamma\gamma}(\partial_{\mu'}a)\tilde{F}^{\mu\nu}+J^{\nu},
\label{eq1}
\\
\quad
(-\partial'^{2}+m^2_{a})a
&=
-\frac{1}{4}g_{a\gamma\gamma}F_{\mu\nu}\tilde{F}^{\mu\nu}.
\label{eq2}
\end{align}
In terms of the electric and magnetic fields, the equations of motion read
\begin{align}
\bm{\nabla}'\cdot \bm{E}
&=
g_{a\gamma\gamma}(\bm{\nabla}'a)\cdot \bm{B}+J^{0},
\label{gauss}
\\
\quad
\partial_{t'}\bm{E}-\bm{\nabla}'\times\bm{B}
&=
g_{a\gamma\gamma}(\bm{B}\partial_{t'}a +\bm{\nabla}'a\times\bm{E})+\bm{J},
\end{align}
and
\begin{align}
(-\eta^{\mu\nu}\partial_{\mu'}\partial_{\nu'}+m^2_{a})a
&=
-g_{a\gamma\gamma}\bm{E}\cdot\bm{B}
\end{align}
with the electric field 
$\bm{E}(t',\bm{x}')=-\partial_{t'}\bm{A}-\bm{\nabla}'\phi$
and the magnetic field
$\bm{B}=\bm{\nabla}'\times\bm{A}$.
We work in the temporal gauge, 
$\phi(t',\bm{x}')=A^{0}(t',\bm{x}')=0$,
together with the Coulomb gauge,
$\bm{\nabla}'\cdot \bm{A}(t', \bm{x}')=0$.
For the field configuration considered below,
$a=a(t', z')$ and 
$\bm{A}(t', z')=(A_{||}(t', z'), A_{\perp}(t', z'),0)$,
we have
$(\bm{\nabla}'a)\cdot\bm{B}=0$.
Note that $z'$ denotes the propagation distance of the photon.
Therefore, for $J^{0}=0$, these gauge conditions are
consistent with the Gauss constraint in Eq.~\eqref{gauss}.

To investigate axion-photon conversion in a periodically varying background field, we decompose the gauge field into a background and a small fluctuation as
$\bm{A} \to 
\bar{\bm{A}}+\bm{A}$,
while the axion field is treated as a fluctuation around
$a=0$. 
We consider the background vector potential
$\bar{\bm{A}}
=\left(0,-\frac{\bar{B}_{0}}{Q'}\sin\left(Q'z'-\Omega' t'\right), 0\right)$
with a constant magnetic field amplitude
$\bar{B}_{0}$, 
and 
$\Omega'$ and $Q'$ 
are the frequency and wave number of the background modulation, respectively.
This choice gives the background electric and magnetic fields
$\bar{\bm{E}}
=
\left(0, -\frac{\Omega'}{Q'}\bar{B}_{0}\cos(Q'z'-\Omega' t'), 0\right)$
and
$\bar{\bm{B}}
=
\left(\bar{B}_{0}\cos(Q'z'-\Omega' t'), 0, 0\right)$,
and requires the background current
$\bar{\bm{J}}$
\begin{align}
\bar{\bm{J}}
=
\left(
1-\frac{\Omega'^2}{Q'^2}
\right)Q'\bar{B}_{0}\sin(Q'z'-\Omega' t')
\end{align}
to satisfy the Maxwell equation.
Keeping only terms linear in the photon and axion fluctuations, Eqs.~\eqref{eq1} and~\eqref{eq2} reduce to
\begin{align}
\left(
\partial^2_{t'}-\partial^2_{z'}+m^2_{\mathrm{a}}
\right)a
&=
g_{a\gamma\gamma}\bar{B}_{0}
\cos\left(Q'z'-\Omega' t'\right)
\left(
\partial_{t'}+\frac{\Omega'}{Q'}\partial_{z'}
\right)A_{||},
\\
\quad
\left(\partial^2_{t'}-\partial^2_{z'}\right)A_{||}&=
-g_{a\gamma\gamma}\bar{B}_{0}
\cos\left(Q'z'-\Omega' t'\right)
\left(
\partial_{t'}+\frac{\Omega'}{Q'}\partial_{z'}
\right)a,
\end{align}
and
\begin{align}
\left(\partial^2_{t'}-\partial^2_{z'}\right)A_{\perp}&=0.
\label{perp}
\end{align}
Equation~\eqref{perp} shows that $A_{\perp}$ is decoupled from the axion field. 
We therefore focus below on the coupled sector $(A_{||}, a)$.

It is useful to transform to a frame in which the background is spatially homogeneous. 
We introduce the coordinates 
$(t, z)$ through
\begin{align}
t= \gamma
\left(t'-\frac{z'}{v} \right), 
\quad
z= 
\gamma
\left(z'-\frac{t'}{v} \right),
\label{Lorentz}
\end{align}
where 
$v = \Omega'/Q'$ 
is the phase velocity, respectively, and
$\gamma=1/\sqrt{1-1/v^2}$ 
is the Lorentz factor for the time-like case 
$v > 1$
in phase space.
Equivalently, the transformation can be written as
\begin{align}
x^{\mu}=\Lambda^{\mu}_{\nu}x'^{\nu},
\quad
\Lambda^{\mu}_{\nu}=
\begin{pmatrix}
\gamma&0&0&-\gamma/v\\
0&1&0&0\\
0&0&1&0\\
-\gamma /v&0&0&\gamma
\end{pmatrix}.
\end{align}
In the Lorentz transformed frame, the coupled field equations become
\begin{align}
\left(
\partial^2_{t}-\partial^2_{z}+m^2_{\mathrm{a}}
\right)a
&=
\frac{\Omega g}{Q'}
\cos\left(\Omega t\right)
\partial_{z} A_{||},
\label{EOM1}
\\
\quad
\left(\partial^2_{t}-\partial^2_{z}\right)A_{||}
&=
-\frac{\Omega g}{Q'}
\cos\left(\Omega t\right)
\partial_{z} a
\label{EOM2}
\end{align}
with the driving frequency
$\Omega=\Omega'/\gamma$
and
$g=g_{a\gamma\gamma}\bar{B}_{0}$.
Under the Lorentz transformation in Eq.~\eqref{Lorentz}, the background electric and magnetic fields become 
$\bar{\bm{\mathcal{E}}}$ and $\bar{\bm{\mathcal{B}}}$ respectively, as follows:
\begin{align}
\bar{\bm{E}}
\to 
\bar{\bm{\mathcal{E}}}
=
\left(
0,-\frac{\Omega}{Q'}\bar B_0\cos(\Omega t),0
\right),
\quad
\bar{\bm{B}}
\to
\bar{\bm{\mathcal{B}}}
=
\boldsymbol{0}.
\end{align}
Thus, the background magnetic field vanishes after the Lorentz transformation, leaving a spatially homogeneous, time-periodic electric field.
In particular, the coupling factor appearing on the right-hand sides of Eqs.~\eqref{EOM1} and~\eqref{EOM2} satisfies
$\frac{\Omega g}{Q'}\cos(\Omega t)
=-g_{a\gamma\gamma}\bar{\mathcal{E}}_y(t)$,
showing that the axion-photon coupling in this frame is governed by the oscillating background electric field.
The crucial advantage of this frame is that the coefficients of the equations are now periodic in time and independent of $z$.
Similar analyses have also been performed in condensed matter systems~\cite{Oka2026, Kanda2026}. 
Translational invariance along the $z$ direction therefore allows us to decompose the fields into momentum modes,
$A_{||}(t,z)
=e^{ik z}\psi_{\gamma}(t)$
and
$a(t,z)=e^{ik z}\psi_{a}(t)$.
Each value of $k$ can consequently be treated independently.

To rewrite the two second-order equations as a first-order system, we introduce the four-component vector 
$|\Psi(t)\rangle
=(\psi_{\gamma}, \psi_{a}, \partial_{t}(\psi_{\gamma}/k), \partial_{t}(\psi_{a}/k))^{T}$.
The equations of motion can then be written in the Schr\"{o}dinger-like form
\begin{align}
i\frac{d}{dt}|\Psi(t)\rangle
=
\mathcal{H}(t)|\Psi(t)\rangle,
\label{schrodingereq}
\end{align}
with
\begin{align}
\mathcal{H}(t)
=
\mathcal{H}_{0}
+\mathcal{V}
\cos\left(\Omega t\right)
\label{hamiltonian}
\end{align}
and the matrices $\mathcal{H}_{0}$ and $\mathcal{V}$ given by
\begin{align}
\mathcal{H}_{0}
=
\begin{pmatrix}
0&0&ik&0\\
0&0&0&ik\\
-ik&0&0&0\\
0&-i\Omega^2_{\mathrm{a}}(k)/k&0&0
\end{pmatrix},
\quad
\mathcal{V}
=
\frac{\Omega g}{Q'}
\begin{pmatrix}
0&0&0&0\\
0&0&0&0\\
0&1&0&0\\
-1&0&0&0
\end{pmatrix},
\end{align}
where 
$\Omega_{\mathrm{a}}(k)
=\sqrt{k^2+m^2_{a}}$.
We emphasize that Eq.~\eqref{schrodingereq} is a convenient first-order representation of the classical field equations.
The matrix 
$\mathcal{H}(t)$
can be regarded as a transfer matrix, which is well-known method in the context of the photonics~\cite{Tuz2010}.
The important property for the following analysis is its periodicity,
$\mathcal{H}(t)=\mathcal{H}(t+2\pi/\Omega)$.
According to the Floquet theorem~\cite{Sambe1973, Shirley1965, Oka2019, Ikeda2025}, 
any solution of a linear differential equation with a time-periodic coefficient matrix can be written as
$|\Psi(t)\rangle=e^{-i \mu t}|\psi(t)\rangle$
with the Floquet state
$|\psi(t)\rangle=|\psi(t+2\pi/\Omega)\rangle$.
Here, the quantity $\mu$ is the Floquet exponent, which is a complex value due to the non-Hermiticity of the above Hamiltonian Eq.~\eqref{hamiltonian}.
The periodic part of the Floquet state
$|\psi(t)\rangle$
can be expanded in harmonics of the driving frequency as
$|\psi(t)\rangle=\sum_{n}e^{-i n \Omega t}|\psi_{n}\rangle$
because of the discrete time periodicity 
$t \sim t+2\pi/\Omega$. 
Substituting these terms into Eq.~\eqref{schrodingereq} yields the following time-independent Floquet eigenvalue equation
\begin{align}
\left(\mathcal{H}_{0}-n\Omega\right)
|\psi_{n}\rangle
+\frac{\mathcal{V}}{2}|\psi_{n+1}\rangle
+\frac{\mathcal{V}}{2}|\psi_{n-1}\rangle
=
\mu|\psi_{n}\rangle.
\label{sambe}
\end{align}
Equation~\eqref{sambe} may be regarded as an eigenvalue problem in an enlarged space labeled by the integer 
$n$, 
often referred to as the Sambe space. 
The index $n$ labels the Floquet replicas generated by absorption or emission of integer multiples of the driving frequency $\Omega$
(for the cases of photons, see Refs.~\cite{Shirley1965, Oka2019, Ikeda2025}).

An important consequence of Floquet periodicity is that the Floquet exponent is defined only modulo $\Omega$. 
Indeed, the transformations
\begin{align}
\mu \to \mu + m\Omega,
\quad
|\psi(t)\rangle \to e^{im\Omega t}|\psi(t)\rangle
\end{align}
with $m \in \mathbb{Z}$ leave the full solution 
$|\Psi(t)\rangle$ unchanged.
Thus, Floquet exponents whose real parts differ by an integer multiple of $\Omega$ represent the same physical Floquet solution.
In analogy with the crystal momentum defined modulo a reciprocal lattice vector in a spatially periodic system, known as the Bloch theorem, one may therefore restrict 
$\mathrm{Re}[\mu]$ 
to an interval of width $\Omega$.
We refer to such an interval as the Floquet Brillouin zone (FBZ).
Extended-zone representation is useful in Sec. III for identifying the resonant crossings responsible for the instability(e.g., see Ref.~\cite{Hummel2011} in Chapter 5).

Since the Floquet exponent can be complex, 
$\mu=\mathrm{Re}[\mu]+i \mathrm{Im}[\mu]$,
the corresponding solution contains the factor
$
e^{-i\mu t}=e^{\operatorname{Im}[\mu] t}
e^{-i\operatorname{Re}[\mu] t}.
$
Therefore, a mode with 
$\mathrm{Im}[\mu] > 0$
grows exponentially with time, whereas a mode with
$\mathrm{Im}[\mu] < 0$
decays exponentially. 
In the following section, we analyze the real and imaginary parts of the Floquet exponent and show how the unstable modes emerge from crossings between Floquet replicas.

\section{Origin of the Floquet instability}
We first investigate the Floquet spectrum numerically and then analytically determine the conditions for the instability in our model by constructing an effective Hamiltonian near the resonant crossings between Floquet replicas.
\begin{figure}[t]
 \centering
\includegraphics[width=0.8\linewidth]{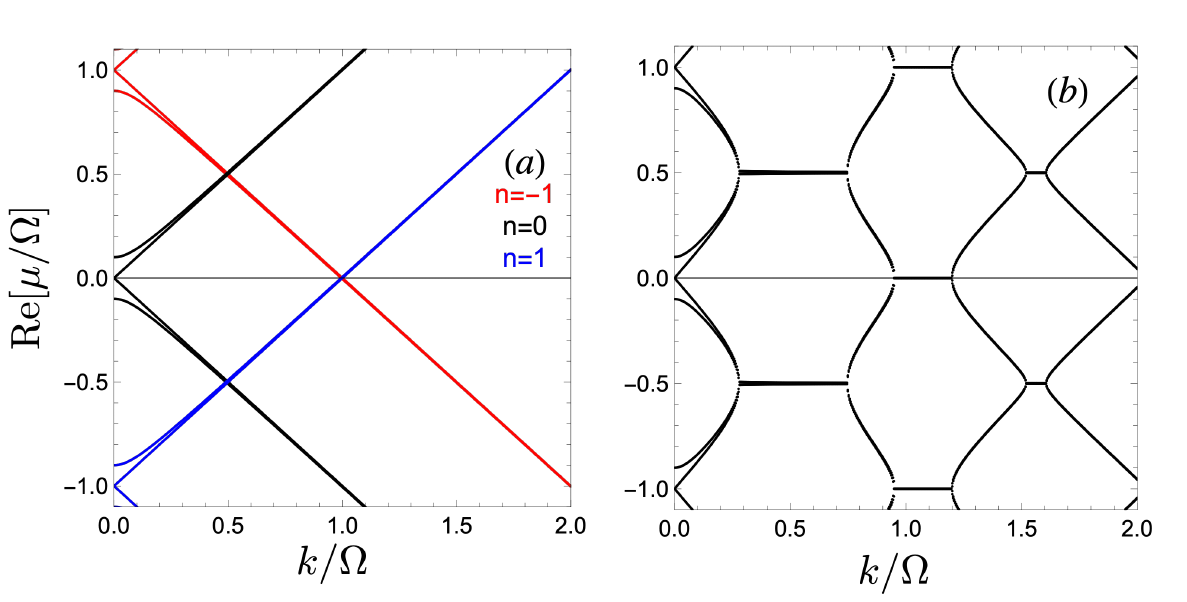}\\
\includegraphics[width=0.8\linewidth]{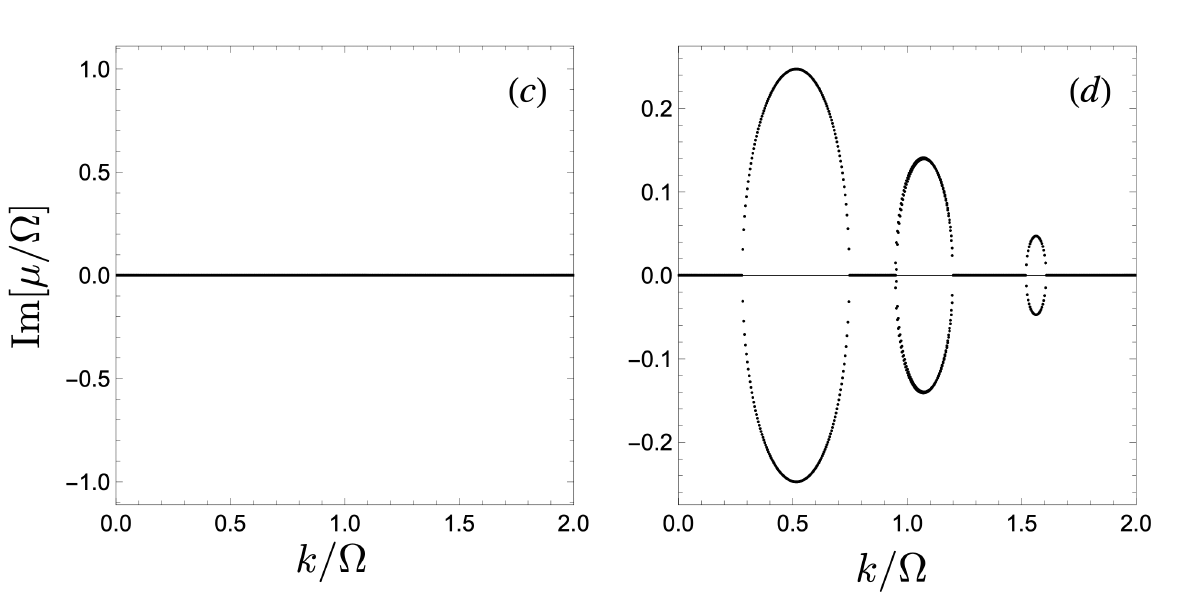}
\caption
    {Real and imaginary parts of the Floquet exponent 
    $\mu$ in the Lorentz-transformed frame. 
    Panels (a) and (b) show 
    $\mathrm{Re}[\mu/\Omega]$, while panels (c) and (d) represent $\mathrm{Im}[\mu/\Omega]$. 
    Panels (a) and (c) correspond to $g/\Omega=0$, and panels (b) and (d) to $g/\Omega=1$. 
    In panel (a), the $n=-1,0,1$ Floquet replicas are indicated by red, black, and blue curves, respectively. 
    We set $m_a/\Omega=0.1$ and $v=1.5$.
    }
\label{fig:imaginarymu}
\end{figure}
Figure~\ref{fig:imaginarymu} shows the real and imaginary parts of the Floquet exponent $\mu$ as functions of the momentum $k$.
Panels (a) and (c) correspond to the uncoupled case
$g/\Omega=0$, whereas panels (b) and (d) correspond to
$g/\Omega=1$.
In the absence of the axion-photon coupling, the Floquet spectrum consists of replicas of the photon and axion branches shifted by integer multiples of the modulation frequency $\Omega$, as shown in Fig.~\ref{fig:imaginarymu}(a).
In this case, all Floquet exponents are real, and hence
$\mathrm{Im}[\mu]=0$, as shown in Fig.~\ref{fig:imaginarymu}(c).
When the axion-photon coupling is switched on, the Floquet replicas hybridize near their crossing points. 
As shown in Fig.~\ref{fig:imaginarymu}(b), some of these crossings are accompanied by the emergence of complex-conjugate pairs of Floquet exponents. 
Correspondingly, a nonzero imaginary part appears in Fig.~\ref{fig:imaginarymu}(d). 
Since a mode behaves as
$|\Psi(t)\rangle \propto e^{\mathrm{Im}[\mu]t}$
a branch with
$\mathrm{Im}[\mu] > 0$
grows exponentially in time and therefore represents an unstable mode induced by Floquet instability.

To understand the origin of the complex Floquet exponents, it is convenient to use the extended-zone representation introduced in Sec. II. 
In the uncoupled limit, the relevant resonance occurs at the crossing between the photon and axion Floquet replicas satisfying
$k+\Omega_{\mathrm{a}}(k)=\Omega$.
The coupling between these resonant modes can be described by an effective Hamiltonian.
The full resonant subspace is four-dimensional, as shown explicitly in Appendix~\ref{A}. 
It decomposes into two independent $2\times2$ blocks. 
One of them is
\begin{align}
\mathcal{H}^{(\mathrm{A})}_{\mathrm{eff}}
=
\begin{pmatrix}
-k+\Omega'&\eta {k}/{\Omega_{\mathrm{a}}}\\
\eta&\Omega_{\mathrm{a}}
\end{pmatrix}
\label{effectivehamiltonian},
\end{align}
where we defined $\eta=i v g/4\gamma$.
The corresponding eigenvalues are
\begin{align}
\mu^{\pm}_{\mathrm{A}}
&=
\frac{1}{2}
\left[
\Omega+\Omega_{\mathrm{a}}-k
\pm \sqrt{(k-\Omega+\Omega_{\mathrm{a}})^2-\frac{g^2 k v^2}{4\gamma^2\Omega_{\mathrm{a}}}}
\right].
\label{EPs}
\end{align}
Equation~\eqref{EPs} directly shows the origin of the instability. 
When 
\begin{align}
(k-\Omega+\Omega_{\mathrm{a}}(k))^2-\frac{g^2 k v^2}{4\gamma^2\Omega_{\mathrm{a}}}<0,
\label{negative}
\end{align}
the square root becomes purely imaginary and the two Floquet exponents form a complex-conjugate pair. 
In the case of which Eq.~\eqref{negative} is satisfied, one of the two modes then has
$\mathrm{Im}[\mu^{+}_{\mathrm{A}}]>0$ and grows exponentially, 
whereas the other has
$\mathrm{Im}[\mu^{-}_{\mathrm{A}}]<0$ and decays exponentially.
The emergence of the instability from resonant crossings between Floquet replicas is consistent with the Floquet-Bloch analysis of axion electrodynamics in Ref.~\cite{Yao20262}. 
In the present model, however, the periodic coupling originates from an external electromagnetic background rather than an oscillating axion background.

The boundaries of the unstable region are exceptional points (EPs), at which the two eigenvalues and the corresponding eigenvectors coalesce. 
They are therefore determined by the vanishing of the discriminant,
\begin{align}
(k-\Omega+\Omega_{\mathrm{a}}(k))^2-\frac{g^2 k v^2}{4\gamma^2\Omega_{\mathrm{a}}}=0.
\label{zero}
\end{align}
A detailed derivation of the effective Hamiltonian and the coalescence of the left and right eigenvectors at the EPs is given in Appendix~\ref{A}.
\begin{figure}[H]
 \centering
\includegraphics[width=0.8\linewidth]{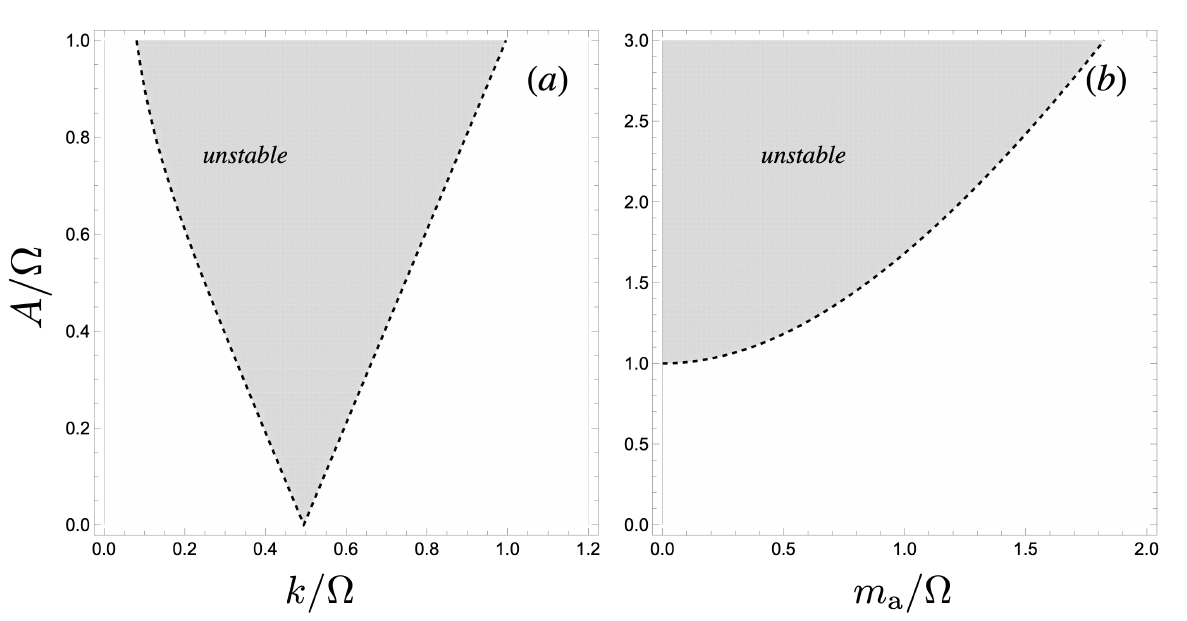}
   \caption
    {Parameter regions for the Floquet instability.
    The shaded regions indicate the parameters for which unstable modes appear, while the solid curves correspond to the exceptional point condition in Eq.~\eqref{negative}. 
    (a) The $(k/\Omega, A/\Omega)$ plane with $m_a/\Omega=0.1$.
    (b) The $(m_a/\Omega, A/\Omega)$ plane with $k/\Omega=1$.}
\label{fig:unstable}
\end{figure}
Fig.~\ref{fig:unstable} 
shows the parameter regions in which unstable modes
appear.
Here, we define $A:=gv/(2\gamma)$.
The shaded regions correspond to the parameters for which the quantity inside the square root in Eq.~\eqref{EPs} becomes negative, and hence the Floquet exponent acquires a nonzero imaginary part.
In panel (a), we show the unstable region in the
$(k/\Omega, A/\Omega)$ plane for fixed $m_a/\Omega=0.1$.
In panel (b), we represent the corresponding region in the
$(m_a/\Omega, A/\Omega)$ plane for fixed $k/\Omega=1$.
The boundaries of the shaded regions are given by the
exceptional point condition in Eq.~\eqref{negative}.
Similar behaviors have already been known in the context of the non-Hermitian Rabi oscillation model~\cite{Lee2015, Wang2018, Lee2020, Zhou2023, Cen2024, Liu2025}.

In the next section, we examine how this Floquet instability affects the axion-field amplitude generated from an initial photon field.

\section{Instability enhanced field amplitude}
In this section, we examine the time evolution of the axion field generated from an initial photon field.
We introduce the normalized axion-field amplitude
$P_{\gamma \to a}(t)$
as
\begin{align}
P_{\gamma \to a}(t)
=
\frac{|\psi_{\mathrm{a}}(t')|^2}{|\psi_{\gamma}(0)|^2}.
\end{align}
We set the initial conditions as
$\psi_\gamma(0)=1,
\psi_a(0)=0,
\partial_t(\psi_\gamma(0)/k)=i,
\partial_t(\psi_a(0)/k)=0.$
With these initial conditions, the axion field is initially absent, and its subsequent evolution is induced through the coupling to the photon field.
\begin{figure}[H]
 \centering
\includegraphics[width=0.8\linewidth]{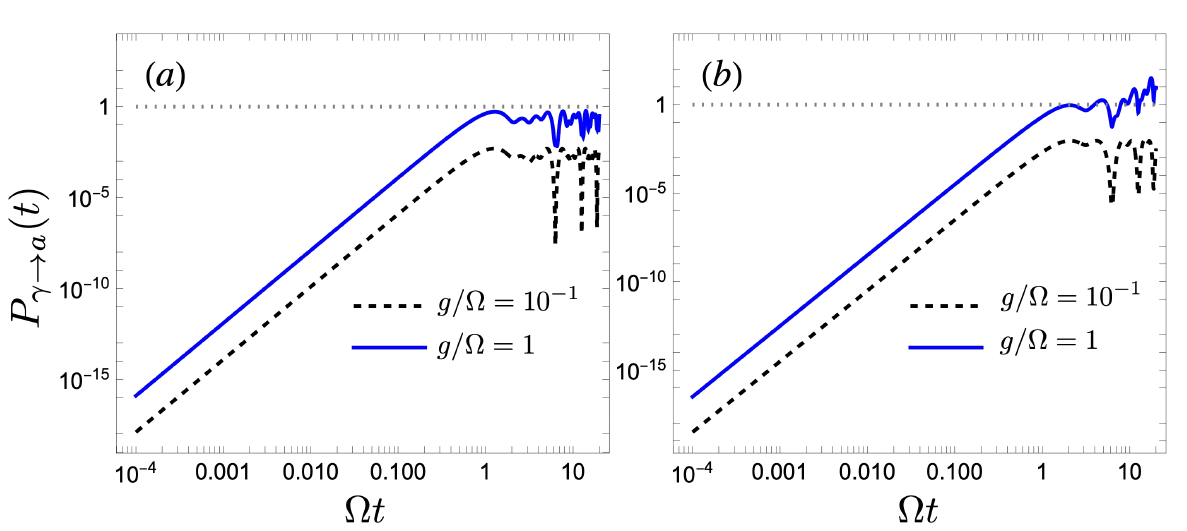}
   \caption
    {Time evolution of the normalized axion-field amplitude $P_{\gamma\to a}(t)$.
    The black dashed and blue solid curves correspond to
    $g/\Omega=10^{-1}$ and $g/\Omega=1$, respectively.
    Panels (a) and (b) show the results for
    $k/\Omega=2$ and $k/\Omega=1$, respectively.
    We set $m_a/\Omega=0.1$ and $v=1.5$.
    The initial conditions are
    $\psi_\gamma(0)=1$, $\psi_a(0)=0$,
    $\partial_t(\psi_\gamma(0)/k)=i$, and
    $\partial_t(\psi_a(0)/k)=0$.
    }
\label{fig:transitionp}
\end{figure}
Figure~\ref{fig:transitionp} shows the time evolution of the axion-field amplitude
$P_{\gamma\to a}(t)$.
Panels (a) and (b) correspond to 
$k/\Omega=2$ and $k/\Omega=1$, respectively.
As shown in Fig.~\ref{fig:imaginarymu}(d), the mode around 
$k/\Omega=1$
enters the unstable region for the strong coupling
$g/\Omega=1$, 
whereas the mode at $k/\Omega=2$ remains stable.

In panel (a), the axion-field amplitude remains bounded even for the stronger coupling.
In contrast, in panel (b), the axion-field amplitude is strongly enhanced for 
$g/\Omega=1$, 
where the unstable mode is present. 
This comparison demonstrates that the Floquet instability induced by the periodic axion-photon coupling enhances the axion field generated from the initial photon field.

\section{Conclusion}
The axion is one of the theoretically well-motivated candidates for physics beyond the Standard Model.
Axion-photon conversion provides an important avenue for probing the existence and properties of axions.
In this paper, we investigated Floquet instability in axion-photon conversion in a simplified model motivated by photons orbiting a black hole. 
We considered a periodically varying electromagnetic background that induces a time-periodic axion-photon coupling after a Lorentz transformation.
By analyzing the Floquet spectrum, we found unstable modes associated with resonant crossings between Floquet replicas, consistent with the general Floquet-Bloch picture discussed in Ref.~\cite{Yao20262}.
We determined the parameter regions in which the instability occurs by deriving the corresponding exceptional point conditions.
We further showed that the presence of the Floquet instability enhances the axion-field amplitude generated from an initial photon field.
Our results show that a periodic electromagnetic coupling can significantly affect axion-photon conversion in a setting motivated by photon motion around a black hole.

As a future direction, the present mechanism identified here may offer a new perspective for axion searches.
Although the direct detection of axions in realistic astrophysical environments, such as black hole systems, remains challenging, our analysis may provide useful insights toward this broader goal.
To pursue this direction, we should compare the parameter regions where the instability occurs with existing observational constraints. 
More detailed quantitative assessments of this comparison are left for future work.
An alternative direction is to investigate an analog realization of the Floquet instability studied here in condensed matter systems.
In these systems, the relevant parameters can be controlled more flexibly than in astrophysical or cosmological settings~\cite{Unruh1981, Unruh2005, Jacquet2020, Hotta2022, Tsuda2023, Sedrakian2024, Katayama2025, Barcelo2026}.
These analog platforms could enable table-top experiments to study the Floquet instability mechanism responsible for enhancing axion-photon conversion.
More broadly, this direction could help bridge cosmology and condensed matter physics.

\acknowledgments
We thank Jiro Soda for useful discussions. 
We also thank R.-M. Yao for bringing Ref.~\cite{Yao20262} to our attention.
Y.S. is supported by JSPS KAKENHI Grant Number JP25KJ0065. 
K.U. is supported by JSPS KAKENHI Grant Number JP24K17050.

\begin{appendix}
\section{Perturbative analysis of the Floquet Hamiltonian Eq.~\eqref{effectivehamiltonian}
\label{A}}
In this appendix, we perturbatively derive an effective Hamiltonian in Eq.~\eqref{effectivehamiltonian} used in Sec.~III.
Then, we analytically determine the boundaries where the unstable region induced by Floquet instability appears.

We first diagonalize the unperturbed part $\mathcal{H}_0$.
Since $\mathcal{H}_0$ 
is in general non-Hermitian, it is
diagonalized by a similarity transformation rather than a unitary
transformation
$\mathcal{H}_{0}
=S\tilde{\mathcal{H}}_{0}S^{-1}$, where the columns of $S$ are the right eigenvectors of
$\mathcal{H}_0$.
The transformation matrix $S$ is given by
\begin{align}
S=
\begin{pmatrix}
i&-i&0&0\\
0&0&ik/\Omega_{\mathrm{a}}&-ik/\Omega_{\mathrm{a}}\\
1&1&0&0\\
0&0&1&1
\end{pmatrix}.
\end{align}
The transformed Hamiltonian
$\tilde{\mathcal{H}}_{0}$
is then in diagonal form
$\tilde{\mathcal{H}}_{0}
=\mathrm{diag}
(k, -k, \Omega_{\mathrm{a}}, -\Omega_{\mathrm{a}})$.
Introducing the transformed state
$|\Phi(t)\rangle=S^{-1}|\Psi(t)\rangle$,
the Schr\"{o}dinger-like equation~\eqref{schrodingereq} becomes
\begin{align}
i\frac{d}{dt}
|\Phi(t)\rangle
=
\left(
\tilde{\mathcal{H}}_{0}
+\tilde{\mathcal{V}}\cos(\Omega t)
\right)
|\Phi(t)\rangle.
\label{transformschro}
\end{align}
Here, the transformed interaction matrix
$\tilde{\mathcal{V}}$ is defined by
\begin{align}
\tilde{\mathcal{V}}
=
S^{-1}\mathcal{V}S
=
\frac{i v g}{2\gamma}
\begin{pmatrix}
0&0&k/\Omega_{\mathrm{a}}&-k/\Omega_{\mathrm{a}}\\
0&0&k/\Omega_{\mathrm{a}}&-k/\Omega_{\mathrm{a}}\\
-1&1&0&0\\
-1&1&0&0
\end{pmatrix}.
\end{align}
Since the transformed Hamiltonian 
$S^{-1} \mathcal{H}(t) S$
is also periodic in time with period
$2\pi/\Omega$, we apply the Floquet ansatz
$|\Phi(t)\rangle
=e^{-i \mu t}|\phi(t)\rangle$
with the Floquet state
$|\phi(t)\rangle
=\sum_{n}e^{-i n \Omega t}|\phi_{n}\rangle$.
Substituting this expression into Eq.~\eqref{transformschro} yields
\begin{align}
\left(\tilde{\mathcal{H}}_{0}-n \Omega\right)|\phi_{n}\rangle
+\frac{\tilde{\mathcal{V}}}{2}|\phi_{n+1}\rangle
+\frac{\tilde{\mathcal{V}}}{2}|\phi_{n-1}\rangle
=
\mu|\phi_{n}\rangle.
\end{align}
Here, the four-dimensional basis in the 
$n$-th Floquet sector $|\phi_{n}\rangle$
is denoted by
$|\phi_{n}\rangle
=(|1,n\rangle, |2,n\rangle, |3,n\rangle, |4,n\rangle)^{T}$.
In the Sambe space representation, this eigenvalue equation takes the block tridiagonal form
\begin{align}
\begin{pmatrix}
\ddots & \ddots & 0      & 0      & 0\\
\ddots & \tilde{\mathcal{H}}_{0}+\Omega& \tilde{\mathcal{V}}/2 & 0 & 0\\
0 & \tilde{\mathcal{V}}/2 & \tilde{\mathcal{H}}_{0} & \tilde{\mathcal{V}}/2 & 0 \\
0 & 0 & \tilde{\mathcal{V}}/2& \tilde{\mathcal{H}}_{0}-\Omega
& \ddots \\
0 & 0   & 0   & \ddots & \ddots
\end{pmatrix}.
\end{align}

As an example, we now construct an effective Hamiltonian near the modes between
$n=-1$ and $n=0$ Floquet sectors.
The corresponding part of the Floquet Hamiltonian consists of 
$8\times8$ matrix
and is written as
\begin{align}
\begin{pmatrix}
\tilde{\mathcal{H}}_{0}+\Omega&\tilde{\mathcal{V}}/2\\
\tilde{\mathcal{V}}/2&\tilde{\mathcal{H}}_{0}
\end{pmatrix}.
\label{A6}
\end{align}
In the absence of the axion-photon coupling, $g=0$, 
the unperturbed Floquet exponents (quasi-energies) in these two sectors are

\begin{align}
\tilde{\mathcal{H}}_{0}
|\phi_{0}\rangle
=E_{j,0}|\phi_{0}\rangle,
\quad
(\tilde{\mathcal{H}}_{0}+\Omega)
|\phi_{-1}\rangle
=E_{j,-1}|\phi_{-1}\rangle
\label{A7}
\end{align}
with
\begin{align}
E_{j,0}
&=
\left(
k,-k,\Omega_{\mathrm{a}},-\Omega_{\mathrm{a}}
\right),
\\
E_{j,-1}
&=
\left(
k+\Omega,-k+\Omega,
\Omega_{\mathrm{a}}+\Omega,
-\Omega_{\mathrm{a}}+\Omega
\right).
\end{align}
Since the matrix in Eq.~\eqref{A6} takes the diagonal form for 
$g=0$ case, 
$E_{j,0}$ and $E_{j, -1}$
correspond to its eigenvalues in Eq.\eqref{A7}.
In this case, the Floquet replicas exhibit several distinct band crossings between the 
$n=-1$ and $n=0$ sectors.
We focus on the crossings, which is shown in Fig. 2(a) in the main text satisfying
\begin{align}
E_{1,0}=E_{4,-1}, 
\quad
E_{3,0}=E_{2,-1},
\end{align}
both of which occur at
\begin{align}
k+\Omega_{\mathrm{a}}(k)=\Omega.
\label{crossings}
\end{align}
This condition defines the crossings in Fig. 2(a).

We next examine the effect of a finite axion-photon coupling $g$ on these degeneracies.
For this purpose, we consider the four states involved in the two crossings
$\{|2, -1\rangle, |4, -1\rangle, |1,0\rangle, |3, 0\rangle\}$.
In this basis, the effective Hamiltonian is given by
\begin{align}
\mathcal{H}^{(4)}_{\mathrm{eff}}
=
\begin{pmatrix}
-k+\Omega&0&0&\eta {k}/{\Omega_{\mathrm{a}}}\\
0&-\Omega_{\mathrm{a}}+\Omega&-\eta&0\\
0&-\eta {k}/{\Omega_{\mathrm{a}}}&k&0\\
\eta&0&0&\Omega_{\mathrm{a}}
\end{pmatrix}
\end{align}
with
$\eta
={i v g}/{4\gamma}$.
The four-dimensional effective Hamiltonian can be decomposed into two independent $2\times2$ blocks.
By rearranging the order of the basis from
$\{|2, -1\rangle, |4, -1\rangle, |1,0\rangle, |3, 0\rangle\}$
to
$\{|2, -1\rangle, |3, 0\rangle, |4, -1\rangle, |1,0\rangle\}$,
the effective Hamiltonian takes the block diagonal form (e.g., ~\cite{Wang2015})
\begin{align}
\mathcal{H}^{(4)}_{\mathrm{eff}}
=
\mathcal{H}^{(\mathrm{A})}_{\mathrm{eff}}
\oplus
\mathcal{H}^{(\mathrm{B})}_{\mathrm{eff}},
\end{align}
where
$\mathcal{H}^{(\mathrm{A})}_{\mathrm{eff}}$
and
$\mathcal{H}^{(\mathrm{B})}_{\mathrm{eff}}$
are both $2\times2$ matrices introduced as
\begin{align}
\mathcal{H}^{(\mathrm{A})}_{\mathrm{eff}}
=
\begin{pmatrix}
-k+\Omega&\eta {k}/{\Omega_{\mathrm{a}}}\\
\eta&\Omega_{\mathrm{a}}
\end{pmatrix},
\quad
\mathcal{H}^{(\mathrm{B})}_{\mathrm{eff}}
=
\begin{pmatrix}
-\Omega_{\mathrm{a}}+\Omega&-\eta\\
-\eta {k}/{\Omega_{\mathrm{a}}}&k
\end{pmatrix}.
\end{align}
The eigenvalues of 
$\mathcal{H}^{(\mathrm{A})}_{\mathrm{eff}}$ 
and
$\mathcal{H}^{(\mathrm{B})}_{\mathrm{eff}}$
are given by
\begin{align}
\mu^{\pm}_{\mathrm{A}}
&=
\frac{1}{2}
\left[
\Omega+\Omega_{\mathrm{a}}-k
\pm \sqrt{(k-\Omega+\Omega_{\mathrm{a}})^2-\frac{g^2 k v^2}{4\gamma^2\Omega_{\mathrm{a}}}}
\right],
\label{evA}
\\
\quad
\mu^{\pm}_{\mathrm{B}}
&=
\frac{1}{2}
\left[
\Omega-\Omega_{\mathrm{a}}+k
\pm \sqrt{(k-\Omega+\Omega_{\mathrm{a}})^2-\frac{g^2 k v^2}{4\gamma^2\Omega_{\mathrm{a}}}}
\right].
\label{evB}
\end{align}

The degeneracy condition for the eigenvalues is obtained by requiring the discriminant in Eqs.~\eqref{evA} and~\eqref{evB} to vanish:
\begin{align}
(k-\Omega+\Omega_{\mathrm{a}}(k))^2-\frac{g^2 k v^2}{4\gamma^2\Omega_{\mathrm{a}}}=0.
\label{discriminant}
\end{align}
It corresponds to Eq.~\eqref{crossings}.
Solving Eq.~\eqref{discriminant} with respect to $k$ determines the momenta at which the eigenvalues become degenerate.
We show below that the corresponding eigenvectors also coalesce, confirming that these points are EPs.

In the weak-coupling regime, $g/\Omega\ll1$, the positions of the EPs can be obtained perturbatively.
For $g=0$, Eq.~(A16) reduces to the resonance condition
$k-\Omega+\Omega_{\mathrm a}(k)=0$.
The corresponding solution $k=k_0$ is
\begin{align}
k_{0}=\frac{\Omega}{2}
\Big[
1-
\left(
\frac{m_{\mathrm{a}}}{\Omega}
\right)^2
\Big].
\end{align}
For finite but small $g$, we write the EP positions as
$k=k_{0}+\delta k$
with 
$\delta k/k_{0} \ll 1$
and  
$\delta k \sim \mathcal{O}(g)$.
Expanding Eq.~\eqref{discriminant} to leading order in $g$, we obtain
\begin{align}
\delta k
&=
\pm
\frac{g v}{2\gamma}
\frac{\sqrt{k_{0}/\Omega^{0}_{\mathrm{a}}}}{1+k_{0}/\Omega^{0}_{\mathrm{a}}}
\nonumber\\
\quad
&=
\pm
\frac{g v}{4\gamma}
\sqrt{1-\left(\frac{m_{\mathrm{a}}}{\Omega}\right)^4},
\label{deltak}
\end{align}
where we defined
$\Omega^{0}_{\mathrm{a}}=\sqrt{k^2_{0}+m^2_{\mathrm{a}}}$.
In the second equality of Eq.~\eqref{deltak}, we used the resonance condition
$k_0-\Omega+\Omega_{\mathrm a}^{0}=0$.
Thus, in the weak-coupling regime, the two EPs are located at
$k=k_0\pm|\delta k|$ 
around the resonance momentum $k_0$.

We finally confirm that the solutions of Eq.~\eqref{discriminant} are EPs by examining the eigenvectors of the effective Hamiltonians
$\mathcal{H}^{\mathrm{(A)}}_{\mathrm{eff}}$ and $\mathcal{H}^{\mathrm{(B)}}_{\mathrm{eff}}$.
The right eigenvectors of
$\mathcal{H}^{(\mathrm{A})}_{\mathrm{eff}}$
and
$\mathcal{H}^{(\mathrm{B})}_{\mathrm{eff}}$
are defined by
\begin{align}
\mathcal{H}^{(\mathrm{A})}_{\mathrm{eff}}
|u^{\mathrm{R}}_{\pm}\rangle
&=
\mu^{\pm}_{\mathrm{A}}|u^{\mathrm{R}}_{\pm}\rangle,
\\
\mathcal{H}^{(\mathrm{B})}_{\mathrm{eff}}
|v^{\mathrm{R}}_{\pm}\rangle
&=
\mu^{\pm}_{\mathrm{B}}|v^{\mathrm{R}}_{\pm}\rangle.
\end{align}
The corresponding right eigenvectors can be written as
\begin{align}
|u^{\mathrm{R}}_{\pm}\rangle
=
\begin{pmatrix}
\dfrac{2\gamma i}{gv}
\left[
\Omega_{\mathrm a}-\Omega+k
\mp
\sqrt{
\left(k-\Omega+\Omega_{\mathrm a}\right)^2
-\dfrac{g^2kv^2}{4\gamma^2\Omega_{\mathrm a}}
}
\right]
\\[2mm]
1
\end{pmatrix},
\end{align}
and
\begin{align}
|v^{\mathrm{R}}_{\pm}\rangle
=
\begin{pmatrix}
-\dfrac{2\gamma\Omega_{\mathrm a}i}{gvk}
\left[
\Omega_{\mathrm a}-\Omega+k
\mp
\sqrt{
\left(k-\Omega+\Omega_{\mathrm a}\right)^2
-\dfrac{g^2kv^2}{4\gamma^2\Omega_{\mathrm a}}
}
\right]
\\[2mm]
1
\end{pmatrix}.
\end{align}
At the EPs, the square-root term vanishes, and the two right eigenvectors of each effective Hamiltonian degenerate:
\begin{align}
|u^{\mathrm{R}}\rangle
:=
|u^{\mathrm{R}}_{+}\rangle
=
|u^{\mathrm{R}}_{-}\rangle
=
\begin{pmatrix}
\dfrac{2\gamma i}{gv}
\left(\Omega_{\mathrm a}-\Omega+k\right)
\\[2mm]
1
\end{pmatrix},
\end{align}
and
\begin{align}
|v^{\mathrm{R}}\rangle
:=
|v^{\mathrm{R}}_{+}\rangle
=
|v^{\mathrm{R}}_{-}\rangle
=
\begin{pmatrix}
-\dfrac{2\gamma\Omega_{\mathrm a}i}{gvk}
\left(\Omega_{\mathrm a}-\Omega+k\right)
\\[2mm]
1
\end{pmatrix}.
\end{align}
The corresponding left eigenvectors at the EPs are defined by
\begin{align}
\langle u^{\mathrm{L}}|
\mathcal{H}^{(\mathrm{A})}_{\mathrm{eff}}
&=
\mu_{\mathrm{A}}\langle u^{\mathrm{L}}|,
\\
\langle v^{\mathrm{L}}|
\mathcal{H}^{(\mathrm{B})}_{\mathrm{eff}}
&=
\mu_{\mathrm{B}}\langle v^{\mathrm{L}}|,
\end{align}
and are given by
\begin{align}
\langle u^{\mathrm{L}}|
&=
\begin{pmatrix}
\dfrac{ivg}
{2\gamma(\Omega_{\mathrm a}-\Omega+k)}
&
1
\end{pmatrix},
\\
\langle v^{\mathrm{L}}|
&=
\begin{pmatrix}
-\dfrac{ivgk}
{2\gamma\Omega_{\mathrm a}(\Omega_{\mathrm a}-\Omega+k)}
&
1
\end{pmatrix}.
\end{align}
These eigenvectors satisfy the self-orthogonality relations
\begin{align}
\langle u^{\mathrm{L}}|u^{\mathrm{R}}\rangle
=
\langle v^{\mathrm{L}}|v^{\mathrm{R}}\rangle
=
0.
\end{align}
Thus, the eigenvalues and the corresponding eigenvectors coalesce
at the solutions of Eq.~\eqref{discriminant}, confirming that these points are EPs.
\end{appendix}

\bibliographystyle{apsrev4-2}
\bibliography{ref}

\begin{thebibliography}{50}%
\makeatletter
\providecommand \@ifxundefined [1]{%
 \@ifx{#1\undefined}
}%
\providecommand \@ifnum [1]{%
 \ifnum #1\expandafter \@firstoftwo
 \else \expandafter \@secondoftwo
 \fi
}%
\providecommand \@ifx [1]{%
 \ifx #1\expandafter \@firstoftwo
 \else \expandafter \@secondoftwo
 \fi
}%
\providecommand \natexlab [1]{#1}%
\providecommand \enquote  [1]{``#1''}%
\providecommand \bibnamefont  [1]{#1}%
\providecommand \bibfnamefont [1]{#1}%
\providecommand \citenamefont [1]{#1}%
\providecommand \href@noop [0]{\@secondoftwo}%
\providecommand \href [0]{\begingroup \@sanitize@url \@href}%
\providecommand \@href[1]{\@@startlink{#1}\@@href}%
\providecommand \@@href[1]{\endgroup#1\@@endlink}%
\providecommand \@sanitize@url [0]{\catcode `\\12\catcode `\$12\catcode `\&12\catcode `\#12\catcode `\^12\catcode `\_12\catcode `\%12\relax}%
\providecommand \@@startlink[1]{}%
\providecommand \@@endlink[0]{}%
\providecommand \url  [0]{\begingroup\@sanitize@url \@url }%
\providecommand \@url [1]{\endgroup\@href {#1}{\urlprefix }}%
\providecommand \urlprefix  [0]{URL }%
\providecommand \Eprint [0]{\href }%
\providecommand \doibase [0]{https://doi.org/}%
\providecommand \selectlanguage [0]{\@gobble}%
\providecommand \bibinfo  [0]{\@secondoftwo}%
\providecommand \bibfield  [0]{\@secondoftwo}%
\providecommand \translation [1]{[#1]}%
\providecommand \BibitemOpen [0]{}%
\providecommand \bibitemStop [0]{}%
\providecommand \bibitemNoStop [0]{.\EOS\space}%
\providecommand \EOS [0]{\spacefactor3000\relax}%
\providecommand \BibitemShut  [1]{\csname bibitem#1\endcsname}%
\let\auto@bib@innerbib\@empty
\bibitem [{\citenamefont {Peccei}\ and\ \citenamefont {Quinn}(1977{\natexlab{a}})}]{PQ1977}%
  \BibitemOpen
  \bibfield  {author} {\bibinfo {author} {\bibfnamefont {R.~D.}\ \bibnamefont {Peccei}}\ and\ \bibinfo {author} {\bibfnamefont {H.~R.}\ \bibnamefont {Quinn}},\ }\href {https://doi.org/10.1103/PhysRevLett.38.1440} {\bibfield  {journal} {\bibinfo  {journal} {Phys. Rev. Lett.}\ }\textbf {\bibinfo {volume} {38}},\ \bibinfo {pages} {1440} (\bibinfo {year} {1977}{\natexlab{a}})}\BibitemShut {NoStop}%
\bibitem [{\citenamefont {Peccei}\ and\ \citenamefont {Quinn}(1977{\natexlab{b}})}]{PQ19772}%
  \BibitemOpen
  \bibfield  {author} {\bibinfo {author} {\bibfnamefont {R.~D.}\ \bibnamefont {Peccei}}\ and\ \bibinfo {author} {\bibfnamefont {H.~R.}\ \bibnamefont {Quinn}},\ }\href {https://doi.org/10.1103/PhysRevD.16.1791} {\bibfield  {journal} {\bibinfo  {journal} {Phys. Rev. D}\ }\textbf {\bibinfo {volume} {16}},\ \bibinfo {pages} {1791} (\bibinfo {year} {1977}{\natexlab{b}})}\BibitemShut {NoStop}%
\bibitem [{\citenamefont {Weinberg}(1978)}]{Weinberg1978}%
  \BibitemOpen
  \bibfield  {author} {\bibinfo {author} {\bibfnamefont {S.}~\bibnamefont {Weinberg}},\ }\href {https://doi.org/10.1103/PhysRevLett.40.223} {\bibfield  {journal} {\bibinfo  {journal} {Phys. Rev. Lett.}\ }\textbf {\bibinfo {volume} {40}},\ \bibinfo {pages} {223} (\bibinfo {year} {1978})}\BibitemShut {NoStop}%
\bibitem [{\citenamefont {Wilczek}(1978)}]{Wilczek1978}%
  \BibitemOpen
  \bibfield  {author} {\bibinfo {author} {\bibfnamefont {F.}~\bibnamefont {Wilczek}},\ }\href {https://doi.org/10.1103/PhysRevLett.40.279} {\bibfield  {journal} {\bibinfo  {journal} {Phys. Rev. Lett.}\ }\textbf {\bibinfo {volume} {40}},\ \bibinfo {pages} {279} (\bibinfo {year} {1978})}\BibitemShut {NoStop}%
\bibitem [{\citenamefont {O'Hare}(2020)}]{AxionLimits2020}%
  \BibitemOpen
  \bibfield  {author} {\bibinfo {author} {\bibfnamefont {C.}~\bibnamefont {O'Hare}},\ }\href {https://doi.org/10.5281/zenodo.3932430} {\bibinfo {title} {cajohare/axionlimits: Axionlimits}},\ \bibinfo {howpublished} {\url{https://cajohare.github.io/AxionLimits/}} (\bibinfo {year} {2020})\BibitemShut {NoStop}%
\bibitem [{\citenamefont {Maiani}\ \emph {et~al.}(1986)\citenamefont {Maiani}, \citenamefont {Petronzio},\ and\ \citenamefont {Zavattini}}]{Maiani1986}%
  \BibitemOpen
  \bibfield  {author} {\bibinfo {author} {\bibfnamefont {L.}~\bibnamefont {Maiani}}, \bibinfo {author} {\bibfnamefont {R.}~\bibnamefont {Petronzio}},\ and\ \bibinfo {author} {\bibfnamefont {E.}~\bibnamefont {Zavattini}},\ }\href {https://doi.org/https://doi.org/10.1016/0370-2693(86)90869-5} {\bibfield  {journal} {\bibinfo  {journal} {Physics Letters B}\ }\textbf {\bibinfo {volume} {175}},\ \bibinfo {pages} {359} (\bibinfo {year} {1986})}\BibitemShut {NoStop}%
\bibitem [{\citenamefont {Raffelt}\ and\ \citenamefont {Stodolsky}(1988)}]{Raffelt1988}%
  \BibitemOpen
  \bibfield  {author} {\bibinfo {author} {\bibfnamefont {G.}~\bibnamefont {Raffelt}}\ and\ \bibinfo {author} {\bibfnamefont {L.}~\bibnamefont {Stodolsky}},\ }\href {https://doi.org/10.1103/PhysRevD.37.1237} {\bibfield  {journal} {\bibinfo  {journal} {Phys. Rev. D}\ }\textbf {\bibinfo {volume} {37}},\ \bibinfo {pages} {1237} (\bibinfo {year} {1988})}\BibitemShut {NoStop}%
\bibitem [{\citenamefont {Ringwald}(2024)}]{Ringwald2024}%
  \BibitemOpen
  \bibfield  {author} {\bibinfo {author} {\bibfnamefont {A.}~\bibnamefont {Ringwald}},\ }\href {https://arxiv.org/abs/2404.09036} {\bibinfo {title} {Review on axions}} (\bibinfo {year} {2024}),\ \Eprint {https://arxiv.org/abs/2404.09036} {arXiv:2404.09036 [hep-ph]} \BibitemShut {NoStop}%
\bibitem [{\citenamefont {Baryakhtar}\ \emph {et~al.}(2025)\citenamefont {Baryakhtar}, \citenamefont {Rosenberg},\ and\ \citenamefont {Rybka}}]{Baryakhtar2025}%
  \BibitemOpen
  \bibfield  {author} {\bibinfo {author} {\bibfnamefont {M.}~\bibnamefont {Baryakhtar}}, \bibinfo {author} {\bibfnamefont {L.}~\bibnamefont {Rosenberg}},\ and\ \bibinfo {author} {\bibfnamefont {G.}~\bibnamefont {Rybka}},\ }\href {https://doi.org/10.1088/1361-6633/ae059d} {\bibfield  {journal} {\bibinfo  {journal} {Reports on Progress in Physics}\ }\textbf {\bibinfo {volume} {88}},\ \bibinfo {pages} {106901} (\bibinfo {year} {2025})}\BibitemShut {NoStop}%
\bibitem [{\citenamefont {Gertsenshtein}(1962)}]{Gertsenshtein1962}%
  \BibitemOpen
  \bibfield  {author} {\bibinfo {author} {\bibfnamefont {M.}~\bibnamefont {Gertsenshtein}},\ }\href {http://jetp.ras.ru/cgi-bin/r/index/r/41/1/p113?a=list} {\bibfield  {journal} {\bibinfo  {journal} {Sov. Phys. JETP}\ }\textbf {\bibinfo {volume} {14}},\ \bibinfo {pages} {84} (\bibinfo {year} {1962})}\BibitemShut {NoStop}%
\bibitem [{\citenamefont {Chen}(1995)}]{Chen1995}%
  \BibitemOpen
  \bibfield  {author} {\bibinfo {author} {\bibfnamefont {P.}~\bibnamefont {Chen}},\ }\href {https://doi.org/10.1103/PhysRevLett.74.634} {\bibfield  {journal} {\bibinfo  {journal} {Phys. Rev. Lett.}\ }\textbf {\bibinfo {volume} {74}},\ \bibinfo {pages} {634} (\bibinfo {year} {1995})}\BibitemShut {NoStop}%
\bibitem [{\citenamefont {Cillis}\ and\ \citenamefont {Harari}(1996)}]{Cillis1996}%
  \BibitemOpen
  \bibfield  {author} {\bibinfo {author} {\bibfnamefont {A.~N.}\ \bibnamefont {Cillis}}\ and\ \bibinfo {author} {\bibfnamefont {D.~D.}\ \bibnamefont {Harari}},\ }\href {https://doi.org/10.1103/PhysRevD.54.4757} {\bibfield  {journal} {\bibinfo  {journal} {Phys. Rev. D}\ }\textbf {\bibinfo {volume} {54}},\ \bibinfo {pages} {4757} (\bibinfo {year} {1996})}\BibitemShut {NoStop}%
\bibitem [{\citenamefont {Kanno}\ \emph {et~al.}(2022)\citenamefont {Kanno}, \citenamefont {Soda},\ and\ \citenamefont {Ueda}}]{Ueda2022}%
  \BibitemOpen
  \bibfield  {author} {\bibinfo {author} {\bibfnamefont {S.}~\bibnamefont {Kanno}}, \bibinfo {author} {\bibfnamefont {J.}~\bibnamefont {Soda}},\ and\ \bibinfo {author} {\bibfnamefont {K.}~\bibnamefont {Ueda}},\ }\href {https://doi.org/10.1103/PhysRevD.106.083508} {\bibfield  {journal} {\bibinfo  {journal} {Phys. Rev. D}\ }\textbf {\bibinfo {volume} {106}},\ \bibinfo {pages} {083508} (\bibinfo {year} {2022})}\BibitemShut {NoStop}%
\bibitem [{\citenamefont {Kanno}\ \emph {et~al.}(2023{\natexlab{a}})\citenamefont {Kanno}, \citenamefont {Mukuno}, \citenamefont {Soda},\ and\ \citenamefont {Ueda}}]{Kanno2023}%
  \BibitemOpen
  \bibfield  {author} {\bibinfo {author} {\bibfnamefont {S.}~\bibnamefont {Kanno}}, \bibinfo {author} {\bibfnamefont {A.}~\bibnamefont {Mukuno}}, \bibinfo {author} {\bibfnamefont {J.}~\bibnamefont {Soda}},\ and\ \bibinfo {author} {\bibfnamefont {K.}~\bibnamefont {Ueda}},\ }\href {https://doi.org/10.1088/1475-7516/2023/05/052} {\bibfield  {journal} {\bibinfo  {journal} {Journal of Cosmology and Astroparticle Physics}\ }\textbf {\bibinfo {volume} {2023}}\bibinfo  {number} { (05)},\ \bibinfo {pages} {052}}\BibitemShut {NoStop}%
\bibitem [{\citenamefont {Kanno}\ \emph {et~al.}(2023{\natexlab{b}})\citenamefont {Kanno}, \citenamefont {Mukuno}, \citenamefont {Soda},\ and\ \citenamefont {Ueda}}]{Kanno20232}%
  \BibitemOpen
\bibfield  {number} {  }\bibfield  {author} {\bibinfo {author} {\bibfnamefont {S.}~\bibnamefont {Kanno}}, \bibinfo {author} {\bibfnamefont {A.}~\bibnamefont {Mukuno}}, \bibinfo {author} {\bibfnamefont {J.}~\bibnamefont {Soda}},\ and\ \bibinfo {author} {\bibfnamefont {K.}~\bibnamefont {Ueda}},\ }\bibfield  {journal} {\bibinfo  {journal} {Physical Review D}\ }\textbf {\bibinfo {volume} {107}},\ \href {https://doi.org/10.1103/physrevd.107.063503} {10.1103/physrevd.107.063503} (\bibinfo {year} {2023}{\natexlab{b}})\BibitemShut {NoStop}%
\bibitem [{\citenamefont {Ikeda}\ \emph {et~al.}(2025)\citenamefont {Ikeda}, \citenamefont {Kaku}, \citenamefont {Kanno},\ and\ \citenamefont {Soda}}]{TIkeda2025}%
  \BibitemOpen
  \bibfield  {author} {\bibinfo {author} {\bibfnamefont {T.}~\bibnamefont {Ikeda}}, \bibinfo {author} {\bibfnamefont {Y.}~\bibnamefont {Kaku}}, \bibinfo {author} {\bibfnamefont {S.}~\bibnamefont {Kanno}},\ and\ \bibinfo {author} {\bibfnamefont {J.}~\bibnamefont {Soda}},\ }\href {https://doi.org/10.1103/kxs5-yp85} {\bibfield  {journal} {\bibinfo  {journal} {Phys. Rev. D}\ }\textbf {\bibinfo {volume} {112}},\ \bibinfo {pages} {103504} (\bibinfo {year} {2025})}\BibitemShut {NoStop}%
\bibitem [{\citenamefont {Ferreri}(2026)}]{Alessandro2026}%
  \BibitemOpen
  \bibfield  {author} {\bibinfo {author} {\bibfnamefont {A.}~\bibnamefont {Ferreri}},\ }\href {https://arxiv.org/abs/2601.22332} {\bibinfo {title} {Photon-graviton polarization entanglement induced by a classical electromagnetic wave}} (\bibinfo {year} {2026}),\ \Eprint {https://arxiv.org/abs/2601.22332} {arXiv:2601.22332 [gr-qc]} \BibitemShut {NoStop}%
\bibitem [{\citenamefont {Brandenberger}\ \emph {et~al.}(2023)\citenamefont {Brandenberger}, \citenamefont {Delgado}, \citenamefont {Ganz},\ and\ \citenamefont {Lin}}]{Brandenberger2023}%
  \BibitemOpen
  \bibfield  {author} {\bibinfo {author} {\bibfnamefont {R.}~\bibnamefont {Brandenberger}}, \bibinfo {author} {\bibfnamefont {P.~C.}\ \bibnamefont {Delgado}}, \bibinfo {author} {\bibfnamefont {A.}~\bibnamefont {Ganz}},\ and\ \bibinfo {author} {\bibfnamefont {C.}~\bibnamefont {Lin}},\ }\href {https://doi.org/10.1016/j.dark.2023.101202} {\bibfield  {journal} {\bibinfo  {journal} {Physics of the Dark Universe}\ }\textbf {\bibinfo {volume} {40}},\ \bibinfo {pages} {101202} (\bibinfo {year} {2023})}\BibitemShut {NoStop}%
\bibitem [{\citenamefont {Delgado}(2023)}]{Delgado2023}%
  \BibitemOpen
  \bibfield  {author} {\bibinfo {author} {\bibfnamefont {P.~C.~M.}\ \bibnamefont {Delgado}},\ }\href {https://doi.org/10.1103/PhysRevD.108.123539} {\bibfield  {journal} {\bibinfo  {journal} {Phys. Rev. D}\ }\textbf {\bibinfo {volume} {108}},\ \bibinfo {pages} {123539} (\bibinfo {year} {2023})}\BibitemShut {NoStop}%
\bibitem [{\citenamefont {Masaki}\ \emph {et~al.}(2020)\citenamefont {Masaki}, \citenamefont {Aoki},\ and\ \citenamefont {Soda}}]{Masaki2020}%
  \BibitemOpen
  \bibfield  {author} {\bibinfo {author} {\bibfnamefont {E.}~\bibnamefont {Masaki}}, \bibinfo {author} {\bibfnamefont {A.}~\bibnamefont {Aoki}},\ and\ \bibinfo {author} {\bibfnamefont {J.}~\bibnamefont {Soda}},\ }\href {https://doi.org/10.1103/PhysRevD.101.043505} {\bibfield  {journal} {\bibinfo  {journal} {Phys. Rev. D}\ }\textbf {\bibinfo {volume} {101}},\ \bibinfo {pages} {043505} (\bibinfo {year} {2020})}\BibitemShut {NoStop}%
\bibitem [{\citenamefont {Jiao}\ \emph {et~al.}(2025)\citenamefont {Jiao}, \citenamefont {Brandenberger},\ and\ \citenamefont {Kamali}}]{Jiao2025}%
  \BibitemOpen
  \bibfield  {author} {\bibinfo {author} {\bibfnamefont {H.}~\bibnamefont {Jiao}}, \bibinfo {author} {\bibfnamefont {R.}~\bibnamefont {Brandenberger}},\ and\ \bibinfo {author} {\bibfnamefont {V.}~\bibnamefont {Kamali}},\ }\href {https://doi.org/10.1103/9jm1-ss4t} {\bibfield  {journal} {\bibinfo  {journal} {Phys. Rev. D}\ }\textbf {\bibinfo {volume} {112}},\ \bibinfo {pages} {043546} (\bibinfo {year} {2025})}\BibitemShut {NoStop}%
\bibitem [{\citenamefont {Yao}\ \emph {et~al.}(2026{\natexlab{a}})\citenamefont {Yao}, \citenamefont {Bi}, \citenamefont {Yin},\ and\ \citenamefont {Huang}}]{Yao2026}%
  \BibitemOpen
  \bibfield  {author} {\bibinfo {author} {\bibfnamefont {R.-M.}\ \bibnamefont {Yao}}, \bibinfo {author} {\bibfnamefont {X.-J.}\ \bibnamefont {Bi}}, \bibinfo {author} {\bibfnamefont {P.-F.}\ \bibnamefont {Yin}},\ and\ \bibinfo {author} {\bibfnamefont {Q.-G.}\ \bibnamefont {Huang}},\ }\href {https://arxiv.org/abs/2601.02115} {\bibinfo {title} {Resonant photon-axion mixing driven by dark matter oscillations}} (\bibinfo {year} {2026}{\natexlab{a}}),\ \Eprint {https://arxiv.org/abs/2601.02115} {arXiv:2601.02115 [hep-ph]} \BibitemShut {NoStop}%
\bibitem [{\citenamefont {Brandenberger}\ \emph {et~al.}(2026)\citenamefont {Brandenberger}, \citenamefont {Fr{\"o}hlich},\ and\ \citenamefont {Jiao}}]{Brandenberger2026}%
  \BibitemOpen
  \bibfield  {author} {\bibinfo {author} {\bibfnamefont {R.}~\bibnamefont {Brandenberger}}, \bibinfo {author} {\bibfnamefont {J.}~\bibnamefont {Fr{\"o}hlich}},\ and\ \bibinfo {author} {\bibfnamefont {H.}~\bibnamefont {Jiao}},\ }\href {https://doi.org/10.1103/ys32-853g} {\bibfield  {journal} {\bibinfo  {journal} {Phys. Rev. Lett.}\ }\textbf {\bibinfo {volume} {136}},\ \bibinfo {pages} {031001} (\bibinfo {year} {2026})}\BibitemShut {NoStop}%
\bibitem [{\citenamefont {Arias}\ \emph {et~al.}(2016)\citenamefont {Arias}, \citenamefont {Arza},\ and\ \citenamefont {Gamboa}}]{Arias2016}%
  \BibitemOpen
  \bibfield  {author} {\bibinfo {author} {\bibfnamefont {P.}~\bibnamefont {Arias}}, \bibinfo {author} {\bibfnamefont {A.}~\bibnamefont {Arza}},\ and\ \bibinfo {author} {\bibfnamefont {J.}~\bibnamefont {Gamboa}},\ }\bibfield  {journal} {\bibinfo  {journal} {The European Physical Journal C}\ }\textbf {\bibinfo {volume} {76}},\ \href {https://doi.org/10.1140/epjc/s10052-016-4483-0} {10.1140/epjc/s10052-016-4483-0} (\bibinfo {year} {2016})\BibitemShut {NoStop}%
\bibitem [{\citenamefont {Seong}\ \emph {et~al.}(2024)\citenamefont {Seong}, \citenamefont {Sun},\ and\ \citenamefont {Yun}}]{Seong2024}%
  \BibitemOpen
  \bibfield  {author} {\bibinfo {author} {\bibfnamefont {H.}~\bibnamefont {Seong}}, \bibinfo {author} {\bibfnamefont {C.}~\bibnamefont {Sun}},\ and\ \bibinfo {author} {\bibfnamefont {S.}~\bibnamefont {Yun}},\ }\href {https://doi.org/10.1103/PhysRevD.110.015018} {\bibfield  {journal} {\bibinfo  {journal} {Phys. Rev. D}\ }\textbf {\bibinfo {volume} {110}},\ \bibinfo {pages} {015018} (\bibinfo {year} {2024})}\BibitemShut {NoStop}%
\bibitem [{\citenamefont {Yao}\ \emph {et~al.}(2026{\natexlab{b}})\citenamefont {Yao}, \citenamefont {Bi}, \citenamefont {Yin},\ and\ \citenamefont {Huang}}]{Yao20262}%
  \BibitemOpen
  \bibfield  {author} {\bibinfo {author} {\bibfnamefont {R.-M.}\ \bibnamefont {Yao}}, \bibinfo {author} {\bibfnamefont {X.-J.}\ \bibnamefont {Bi}}, \bibinfo {author} {\bibfnamefont {P.-F.}\ \bibnamefont {Yin}},\ and\ \bibinfo {author} {\bibfnamefont {Q.-G.}\ \bibnamefont {Huang}},\ }\href {https://arxiv.org/abs/2607.02232} {\bibinfo {title} {Sideband structure of axion electrodynamics}} (\bibinfo {year} {2026}{\natexlab{b}}),\ \Eprint {https://arxiv.org/abs/2607.02232} {arXiv:2607.02232 [hep-ph]} \BibitemShut {NoStop}%
\bibitem [{\citenamefont {Nomura}\ \emph {et~al.}(2023)\citenamefont {Nomura}, \citenamefont {Saito},\ and\ \citenamefont {Soda}}]{Nomura2023}%
  \BibitemOpen
  \bibfield  {author} {\bibinfo {author} {\bibfnamefont {K.}~\bibnamefont {Nomura}}, \bibinfo {author} {\bibfnamefont {K.}~\bibnamefont {Saito}},\ and\ \bibinfo {author} {\bibfnamefont {J.}~\bibnamefont {Soda}},\ }\bibfield  {journal} {\bibinfo  {journal} {Physical Review D}\ }\textbf {\bibinfo {volume} {107}},\ \href {https://doi.org/10.1103/physrevd.107.123505} {10.1103/physrevd.107.123505} (\bibinfo {year} {2023})\BibitemShut {NoStop}%
\bibitem [{\citenamefont {Oka}(2026)}]{Oka2026}%
  \BibitemOpen
  \bibfield  {author} {\bibinfo {author} {\bibfnamefont {T.}~\bibnamefont {Oka}},\ }\href {https://arxiv.org/abs/2407.21458} {\bibinfo {title} {Shockwave-enhanced floquet engineering of relativistic quasiparticles}} (\bibinfo {year} {2026}),\ \Eprint {https://arxiv.org/abs/2407.21458} {arXiv:2407.21458 [cond-mat.mes-hall]} \BibitemShut {NoStop}%
\bibitem [{\citenamefont {Kanda}\ \emph {et~al.}(2026)\citenamefont {Kanda}, \citenamefont {Fujisawa}, \citenamefont {Yakubo}, \citenamefont {Kawakami},\ and\ \citenamefont {Obuse}}]{Kanda2026}%
  \BibitemOpen
  \bibfield  {author} {\bibinfo {author} {\bibfnamefont {Y.}~\bibnamefont {Kanda}}, \bibinfo {author} {\bibfnamefont {Y.}~\bibnamefont {Fujisawa}}, \bibinfo {author} {\bibfnamefont {K.}~\bibnamefont {Yakubo}}, \bibinfo {author} {\bibfnamefont {N.}~\bibnamefont {Kawakami}},\ and\ \bibinfo {author} {\bibfnamefont {H.}~\bibnamefont {Obuse}},\ }\href {https://arxiv.org/abs/2505.16606} {\bibinfo {title} {Fast and high-fidelity transfer of edge states via dynamical control of topological phases and effects of dissipation}} (\bibinfo {year} {2026}),\ \Eprint {https://arxiv.org/abs/2505.16606} {arXiv:2505.16606 [cond-mat.mes-hall]} \BibitemShut {NoStop}%
\bibitem [{\citenamefont {Tuz}\ \emph {et~al.}(2010)\citenamefont {Tuz}, \citenamefont {Vidil},\ and\ \citenamefont {Prosvirnin}}]{Tuz2010}%
  \BibitemOpen
  \bibfield  {author} {\bibinfo {author} {\bibfnamefont {V.~R.}\ \bibnamefont {Tuz}}, \bibinfo {author} {\bibfnamefont {M.~Y.}\ \bibnamefont {Vidil}},\ and\ \bibinfo {author} {\bibfnamefont {S.~L.}\ \bibnamefont {Prosvirnin}},\ }\href {https://doi.org/10.1088/2040-8978/12/9/095102} {\bibfield  {journal} {\bibinfo  {journal} {Journal of Optics}\ }\textbf {\bibinfo {volume} {12}},\ \bibinfo {pages} {095102} (\bibinfo {year} {2010})}\BibitemShut {NoStop}%
\bibitem [{\citenamefont {Sambe}(1973)}]{Sambe1973}%
  \BibitemOpen
  \bibfield  {author} {\bibinfo {author} {\bibfnamefont {H.}~\bibnamefont {Sambe}},\ }\href {https://doi.org/10.1103/PhysRevA.7.2203} {\bibfield  {journal} {\bibinfo  {journal} {Phys. Rev. A}\ }\textbf {\bibinfo {volume} {7}},\ \bibinfo {pages} {2203} (\bibinfo {year} {1973})}\BibitemShut {NoStop}%
\bibitem [{\citenamefont {Shirley}(1965)}]{Shirley1965}%
  \BibitemOpen
  \bibfield  {author} {\bibinfo {author} {\bibfnamefont {J.~H.}\ \bibnamefont {Shirley}},\ }\href {https://doi.org/10.1103/PhysRev.138.B979} {\bibfield  {journal} {\bibinfo  {journal} {Phys. Rev.}\ }\textbf {\bibinfo {volume} {138}},\ \bibinfo {pages} {B979} (\bibinfo {year} {1965})}\BibitemShut {NoStop}%
\bibitem [{\citenamefont {Oka}\ and\ \citenamefont {Kitamura}(2019)}]{Oka2019}%
  \BibitemOpen
  \bibfield  {author} {\bibinfo {author} {\bibfnamefont {T.}~\bibnamefont {Oka}}\ and\ \bibinfo {author} {\bibfnamefont {S.}~\bibnamefont {Kitamura}},\ }\href {https://doi.org/10.1146/annurev-conmatphys-031218-013423} {\bibfield  {journal} {\bibinfo  {journal} {Annual Review of Condensed Matter Physics}\ }\textbf {\bibinfo {volume} {10}},\ \bibinfo {pages} {387–408} (\bibinfo {year} {2019})}\BibitemShut {NoStop}%
\bibitem [{\citenamefont {Sato}\ and\ \citenamefont {Ikeda}(2025)}]{Ikeda2025}%
  \BibitemOpen
  \bibfield  {author} {\bibinfo {author} {\bibfnamefont {M.}~\bibnamefont {Sato}}\ and\ \bibinfo {author} {\bibfnamefont {T.~N.}\ \bibnamefont {Ikeda}},\ }\bibfield  {journal} {\bibinfo  {journal} {Journal of the Physical Society of Japan}\ }\textbf {\bibinfo {volume} {94}},\ \href {https://doi.org/10.7566/jpsj.94.111007} {10.7566/jpsj.94.111007} (\bibinfo {year} {2025})\BibitemShut {NoStop}%
\bibitem [{\citenamefont {Hummel}(2011)}]{Hummel2011}%
  \BibitemOpen
  \bibfield  {author} {\bibinfo {author} {\bibfnamefont {R.~E.}\ \bibnamefont {Hummel}},\ }\href@noop {} {\emph {\bibinfo {title} {Electronic properties of materials}}}\ (\bibinfo  {publisher} {Springer Science \& Business Media},\ \bibinfo {year} {2011})\BibitemShut {NoStop}%
\bibitem [{\citenamefont {Lee}\ and\ \citenamefont {Joglekar}(2015)}]{Lee2015}%
  \BibitemOpen
  \bibfield  {author} {\bibinfo {author} {\bibfnamefont {T.~E.}\ \bibnamefont {Lee}}\ and\ \bibinfo {author} {\bibfnamefont {Y.~N.}\ \bibnamefont {Joglekar}},\ }\bibfield  {journal} {\bibinfo  {journal} {Physical Review A}\ }\textbf {\bibinfo {volume} {92}},\ \href {https://doi.org/10.1103/physreva.92.042103} {10.1103/physreva.92.042103} (\bibinfo {year} {2015})\BibitemShut {NoStop}%
\bibitem [{\citenamefont {Wang}\ \emph {et~al.}(2018)\citenamefont {Wang}, \citenamefont {Zhang},\ and\ \citenamefont {Chan}}]{Wang2018}%
  \BibitemOpen
  \bibfield  {author} {\bibinfo {author} {\bibfnamefont {N.}~\bibnamefont {Wang}}, \bibinfo {author} {\bibfnamefont {Z.-Q.}\ \bibnamefont {Zhang}},\ and\ \bibinfo {author} {\bibfnamefont {C.~T.}\ \bibnamefont {Chan}},\ }\href {https://doi.org/10.1103/PhysRevB.98.085142} {\bibfield  {journal} {\bibinfo  {journal} {Phys. Rev. B}\ }\textbf {\bibinfo {volume} {98}},\ \bibinfo {pages} {085142} (\bibinfo {year} {2018})}\BibitemShut {NoStop}%
\bibitem [{\citenamefont {Lee}\ and\ \citenamefont {Longhi}(2020)}]{Lee2020}%
  \BibitemOpen
  \bibfield  {author} {\bibinfo {author} {\bibfnamefont {C.~H.}\ \bibnamefont {Lee}}\ and\ \bibinfo {author} {\bibfnamefont {S.}~\bibnamefont {Longhi}},\ }\bibfield  {journal} {\bibinfo  {journal} {Communications Physics}\ }\textbf {\bibinfo {volume} {3}},\ \href {https://doi.org/10.1038/s42005-020-00417-y} {10.1038/s42005-020-00417-y} (\bibinfo {year} {2020})\BibitemShut {NoStop}%
\bibitem [{\citenamefont {Zhou}\ and\ \citenamefont {Zhang}(2023)}]{Zhou2023}%
  \BibitemOpen
  \bibfield  {author} {\bibinfo {author} {\bibfnamefont {L.}~\bibnamefont {Zhou}}\ and\ \bibinfo {author} {\bibfnamefont {D.-J.}\ \bibnamefont {Zhang}},\ }\href {https://doi.org/10.3390/e25101401} {\bibfield  {journal} {\bibinfo  {journal} {Entropy}\ }\textbf {\bibinfo {volume} {25}},\ \bibinfo {pages} {1401} (\bibinfo {year} {2023})}\BibitemShut {NoStop}%
\bibitem [{\citenamefont {Cen}\ \emph {et~al.}(2024)\citenamefont {Cen}, \citenamefont {Joglekar},\ and\ \citenamefont {Saxena}}]{Cen2024}%
  \BibitemOpen
  \bibfield  {author} {\bibinfo {author} {\bibfnamefont {J.}~\bibnamefont {Cen}}, \bibinfo {author} {\bibfnamefont {Y.~N.}\ \bibnamefont {Joglekar}},\ and\ \bibinfo {author} {\bibfnamefont {A.}~\bibnamefont {Saxena}},\ }\href {https://doi.org/10.1103/PhysRevResearch.6.013167} {\bibfield  {journal} {\bibinfo  {journal} {Phys. Rev. Res.}\ }\textbf {\bibinfo {volume} {6}},\ \bibinfo {pages} {013167} (\bibinfo {year} {2024})}\BibitemShut {NoStop}%
\bibitem [{\citenamefont {Liu}\ \emph {et~al.}(2025)\citenamefont {Liu}, \citenamefont {Duan},\ and\ \citenamefont {Chen}}]{Liu2025}%
  \BibitemOpen
  \bibfield  {author} {\bibinfo {author} {\bibfnamefont {Y.}~\bibnamefont {Liu}}, \bibinfo {author} {\bibfnamefont {L.}~\bibnamefont {Duan}},\ and\ \bibinfo {author} {\bibfnamefont {Q.-H.}\ \bibnamefont {Chen}},\ }\href {https://doi.org/10.1103/PhysRevResearch.7.023004} {\bibfield  {journal} {\bibinfo  {journal} {Phys. Rev. Res.}\ }\textbf {\bibinfo {volume} {7}},\ \bibinfo {pages} {023004} (\bibinfo {year} {2025})}\BibitemShut {NoStop}%
\bibitem [{\citenamefont {Unruh}(1981)}]{Unruh1981}%
  \BibitemOpen
  \bibfield  {author} {\bibinfo {author} {\bibfnamefont {W.~G.}\ \bibnamefont {Unruh}},\ }\href {https://doi.org/10.1103/PhysRevLett.46.1351} {\bibfield  {journal} {\bibinfo  {journal} {Phys. Rev. Lett.}\ }\textbf {\bibinfo {volume} {46}},\ \bibinfo {pages} {1351} (\bibinfo {year} {1981})}\BibitemShut {NoStop}%
\bibitem [{\citenamefont {Sch{\"u}tzhold}\ and\ \citenamefont {Unruh}(2005)}]{Unruh2005}%
  \BibitemOpen
  \bibfield  {author} {\bibinfo {author} {\bibfnamefont {R.}~\bibnamefont {Sch{\"u}tzhold}}\ and\ \bibinfo {author} {\bibfnamefont {W.~G.}\ \bibnamefont {Unruh}},\ }\href {https://doi.org/10.1103/PhysRevLett.95.031301} {\bibfield  {journal} {\bibinfo  {journal} {Phys. Rev. Lett.}\ }\textbf {\bibinfo {volume} {95}},\ \bibinfo {pages} {031301} (\bibinfo {year} {2005})}\BibitemShut {NoStop}%
\bibitem [{\citenamefont {Jacquet}\ \emph {et~al.}(2020)\citenamefont {Jacquet}, \citenamefont {Weinfurtner},\ and\ \citenamefont {K{\"o}nig}}]{Jacquet2020}%
  \BibitemOpen
  \bibfield  {author} {\bibinfo {author} {\bibfnamefont {M.~J.}\ \bibnamefont {Jacquet}}, \bibinfo {author} {\bibfnamefont {S.}~\bibnamefont {Weinfurtner}},\ and\ \bibinfo {author} {\bibfnamefont {F.}~\bibnamefont {K{\"o}nig}},\ }\href@noop {} {\bibfield  {journal} {\bibinfo  {journal} {Philosophical transactions. Series A, Mathematical, physical, and engineering sciences}\ }\textbf {\bibinfo {volume} {378}},\ \bibinfo {pages} {20190239} (\bibinfo {year} {2020})}\BibitemShut {NoStop}%
\bibitem [{\citenamefont {Hotta}\ \emph {et~al.}(2022)\citenamefont {Hotta}, \citenamefont {Nambu}, \citenamefont {Sugiyama}, \citenamefont {Yamamoto},\ and\ \citenamefont {Yusa}}]{Hotta2022}%
  \BibitemOpen
  \bibfield  {author} {\bibinfo {author} {\bibfnamefont {M.}~\bibnamefont {Hotta}}, \bibinfo {author} {\bibfnamefont {Y.}~\bibnamefont {Nambu}}, \bibinfo {author} {\bibfnamefont {Y.}~\bibnamefont {Sugiyama}}, \bibinfo {author} {\bibfnamefont {K.}~\bibnamefont {Yamamoto}},\ and\ \bibinfo {author} {\bibfnamefont {G.}~\bibnamefont {Yusa}},\ }\href {https://doi.org/10.1103/PhysRevD.105.105009} {\bibfield  {journal} {\bibinfo  {journal} {Phys. Rev. D}\ }\textbf {\bibinfo {volume} {105}},\ \bibinfo {pages} {105009} (\bibinfo {year} {2022})}\BibitemShut {NoStop}%
\bibitem [{\citenamefont {Tsuda}\ \emph {et~al.}(2023)\citenamefont {Tsuda}, \citenamefont {Tomizawa},\ and\ \citenamefont {Suzuki}}]{Tsuda2023}%
  \BibitemOpen
  \bibfield  {author} {\bibinfo {author} {\bibfnamefont {R.}~\bibnamefont {Tsuda}}, \bibinfo {author} {\bibfnamefont {S.}~\bibnamefont {Tomizawa}},\ and\ \bibinfo {author} {\bibfnamefont {R.}~\bibnamefont {Suzuki}},\ }\href {https://doi.org/10.1103/PhysRevD.107.104020} {\bibfield  {journal} {\bibinfo  {journal} {Phys. Rev. D}\ }\textbf {\bibinfo {volume} {107}},\ \bibinfo {pages} {104020} (\bibinfo {year} {2023})}\BibitemShut {NoStop}%
\bibitem [{\citenamefont {Sedrakian}\ and\ \citenamefont {Rau}(2025)}]{Sedrakian2024}%
  \BibitemOpen
  \bibfield  {author} {\bibinfo {author} {\bibfnamefont {A.}~\bibnamefont {Sedrakian}}\ and\ \bibinfo {author} {\bibfnamefont {P.~B.}\ \bibnamefont {Rau}},\ }\href {https://doi.org/10.1103/PhysRevD.111.023044} {\bibfield  {journal} {\bibinfo  {journal} {Phys. Rev. D}\ }\textbf {\bibinfo {volume} {111}},\ \bibinfo {pages} {023044} (\bibinfo {year} {2025})}\BibitemShut {NoStop}%
\bibitem [{\citenamefont {Katayama}\ and\ \citenamefont {Hatakenaka}(2025)}]{Katayama2025}%
  \BibitemOpen
  \bibfield  {author} {\bibinfo {author} {\bibfnamefont {H.}~\bibnamefont {Katayama}}\ and\ \bibinfo {author} {\bibfnamefont {N.}~\bibnamefont {Hatakenaka}},\ }\href {https://doi.org/10.1103/mn34-7bj5} {\bibfield  {journal} {\bibinfo  {journal} {Phys. Rev. Lett.}\ }\textbf {\bibinfo {volume} {135}},\ \bibinfo {pages} {046001} (\bibinfo {year} {2025})}\BibitemShut {NoStop}%
\bibitem [{\citenamefont {Barcel{\'o}}\ \emph {et~al.}(2026)\citenamefont {Barcel{\'o}}, \citenamefont {Liberati},\ and\ \citenamefont {Visser}}]{Barcelo2026}%
  \BibitemOpen
  \bibfield  {author} {\bibinfo {author} {\bibfnamefont {C.}~\bibnamefont {Barcel{\'o}}}, \bibinfo {author} {\bibfnamefont {S.}~\bibnamefont {Liberati}},\ and\ \bibinfo {author} {\bibfnamefont {M.}~\bibnamefont {Visser}},\ }\href@noop {} {\bibfield  {journal} {\bibinfo  {journal} {Living Reviews in Relativity}\ }\textbf {\bibinfo {volume} {29}},\ \bibinfo {pages} {2} (\bibinfo {year} {2026})}\BibitemShut {NoStop}%
\bibitem [{\citenamefont {Wang}\ \emph {et~al.}(2015)\citenamefont {Wang}, \citenamefont {Liang}, \citenamefont {Yao},\ and\ \citenamefont {Hu}}]{Wang2015}%
  \BibitemOpen
  \bibfield  {author} {\bibinfo {author} {\bibfnamefont {Z.}~\bibnamefont {Wang}}, \bibinfo {author} {\bibfnamefont {Q.-F.}\ \bibnamefont {Liang}}, \bibinfo {author} {\bibfnamefont {D.-X.}\ \bibnamefont {Yao}},\ and\ \bibinfo {author} {\bibfnamefont {X.}~\bibnamefont {Hu}},\ }\bibfield  {journal} {\bibinfo  {journal} {Scientific Reports}\ }\textbf {\bibinfo {volume} {5}},\ \href {https://doi.org/10.1038/srep11686} {10.1038/srep11686} (\bibinfo {year} {2015})\BibitemShut {NoStop}%
\end{thebibliography}%

\end{document}